%% file: main.tex
\documentclass[letterpaper,twocolumn,10pt]{article}
\usepackage{usenix}

\usepackage{tikz}
\usepackage{amsmath}

\usepackage{cite}
\usepackage{amsfonts}
\usepackage{algorithmic}
\usepackage{graphicx}
\usepackage{textcomp}
\usepackage{xcolor}

\usepackage{caption}

\usepackage{xurl}

\usepackage{authblk}

\usepackage{enumitem} 

\newcommand{\hitfilter}{\textit{NoHit}}
\newcommand{\farr}{\textit{FARR}}
\newcommand{\sqinv}{\textit{SFill-Inv}}
\newcommand{\sdr}{\textit{NEWS}}

\newcommand{\cachename}{Speculative and Timing Attack Resilient (\starname) Cache}
\newcommand{\cacheabbrev}{STAR Cache}
\newcommand{\starname}{STAR}
\newcommand{\cachebasic}{NoSFill-FARR-NoHit}
\newcommand{\cacheinvfa}{\starname-FARR}
\newcommand{\cacheinvfaone}{\starname-FARR-T1}
\newcommand{\cacheinvfatwo}{\starname-FARR-T2}
\newcommand{\cacheop}{\starname-NEWS}
\newcommand{\bheading}[1]{{\vspace{2pt}\noindent{\textbf{#1}}\hspace{2pt}}}

\newcommand{\reffig}[1]{Figure~\ref{#1}}
\newcommand{\refsec}[1]{Section~\ref{#1}}
\newcommand{\reftbl}[1]{Table~\ref{#1}}

\newcommand{\opsetup}{\emph{Setup}}
\newcommand{\opauth}{\emph{Authorize}}
\newcommand{\opaccess}{\emph{Access}}
\newcommand{\opuse}{\emph{Use}}
\newcommand{\opsend}{\emph{Send}}
\newcommand{\oprecv}{\emph{Receive}}

\newif\iflongver

\longverfalse

\begin{document}

\date{}

\title{\textbf{Protecting Cache States Against Both Speculative Execution Attacks and Side-channel Attacks}}

\author[]{Guangyuan Hu}
\author[]{Ruby B. Lee}
  \affil[]{Princeton University}
  \affil[]{\textit{\{gh9,rblee\}@princeton.edu}}

\maketitle

\begin{abstract}

Hardware caches are essential performance optimization features in modern processors to reduce the effective memory access time. Unfortunately, they are also the prime targets for attacks on computer processors because they are high-bandwidth and reliable side or covert channels for leaking secrets. Conventional cache timing attacks typically leak secret encryption keys, while recent speculative execution attacks typically leak arbitrary illegally-obtained secrets through cache timing channels. While many hardware defenses have been proposed for each class of attacks, we show that those for conventional (non-speculative) cache timing channels do not work for all speculative execution attacks, and vice versa. We maintain that a cache is not secure unless it can defend against both of these major attack classes.

We propose a new methodology and framework for covering such relatively large attack surfaces to produce a Speculative and Timing Attack Resilient (STAR) cache subsystem. We use this to design two comprehensive secure cache architectures, STAR-FARR and STAR-NEWS, that have very low performance overheads of 5.6\% and 6.8\%, respectively. To the best of our knowledge, these are the first secure cache designs that cover both non-speculative cache side channels and cache-based speculative execution attacks.

Our methodology can be used to compose and check other secure cache designs. It can also be extended to other attack classes and hardware systems. Additionally, we also highlight the intrinsic security and performance benefits of a randomized cache like a real Fully Associative cache with Random Replacement (FARR) and a lower-latency, speculation-aware version (NEWS).

\end{abstract}

\input{0100_intro}

\input{0200_background}
\input{0300_threatmodel}
\input{0400_motivation}
\input{0500_architecture}
\input{0600_impl}

\input{0700_security_eval}
\input{0800_performance_eval}

\section{Conclusion}

The cache state is an important target for side-channel attacks and speculative execution attacks for leaking information. Prior hardware defenses covered either side-channel or speculative attacks but not both. To the best of our knowledge, this is the first paper that clearly shows the dimensions of both attack families, highlighting the danger of same-domain speculative attacks, and proposing comprehensive solutions for both attack families. Our new \cacheinvfa~and \cacheop~cache architectures have low performance overhead of 5.6\% and 6.8\%. Our proposed design methodology can be used to design other secure caches and cover other attack classes and different secure hardware subsystems.

\clearpage

\bibliographystyle{acm}
\bibliography{refs}

\end{document}

%% file: 0100_intro.tex
\section{Introduction}

Timing-based side-channel attacks recover a secret by observing timing differences in accessing a resource shared with the victim.  Caches have large timing differences between a cache hit and a cache miss. They have been frequently exploited to leak secret information. 
In recent speculative execution attacks, cache-based covert channels are also used to leak out secrets obtained during the transient execution of mis-predicted execution paths.

In this paper, we focus on defeating attacks that use the microarchitectural cache states to leak secret information. While there are many other types of microarchitectural states that can be used as timing channels, the cache state is one of the most exploited timing channels because of its ubiquity in modern computers, its clear timing differences, high bandwidth, and state persistence that cannot be rapidly cleared.

Since both non-speculative cache side channels and speculative cache-based attacks are attacks on hardware microarchitecture, a hardware microarchitectural solution is preferred. 
Although many hardware defenses have been proposed to prevent information leakage from the cache state, they addressed only one of the attack families but not both. We provide the first comprehensive secure cache architectures, Speculative and Timing Attack Resilient (STAR) cache that defeat both attack families, with low performance overhead.

We present a methodology to analyze the security of various defense mechanisms. Our critical insight is that in speculative execution attacks, the sender and the receiver can be in the same security domain, thus making security domain based defenses ineffective for these attacks.
On the other hand, conventional side-channel attacks assume that the send and receiver are in different security domains. Our analysis shows that none of side-channel defenses and speculative execution defenses can fully cover the attacks in the other category.

We then propose hardware security features to build a comprehensive defense. We show that a fully-associative cache with a random replacement policy (FARR) can be used to prevent contention-based (missed-based) attacks. Cross-domain hit-based attacks can be mitigated by marking the security domain of (the owner of) a cache line. We identify the same-domain, hit-based speculative execution attacks, and show that they can be defended by a new ``\textit{speculatively fill the cache, but invalidate on squash (\textbf{\sqinv})}'' defense which has lower cost and overhead than other defenses covering speculative cache channels.

We further show that the fully-associative cache can be replaced with a new speculation-aware cache, with the same security profile but lower access latency or power. NewCache \cite{newcache, newcache2015asscc} is a randomized cache with dynamic remapping on each cache miss, proposed to defeat contention based side-channel attacks, and has not been broken for 15 years\cite{randcache:analysis2021sp, fixit2021sp, casa}. We show how a comprehensive defense can be designed to also defeat speculative attacks with a speculation-aware, low-latency randomized cache inspired by NewCache. 

Our key contributions are:

\begin{itemize}[leftmargin=*,noitemsep,topsep=0pt]

    \item 
    Showing the attack space covering cache side-channel attacks and cache-based speculative execution attacks. Identifying speculative attacks where the sender and receiver of covert channel are in the same security domain. Showing no existing defense has fully addressed both attack classes.
    
    \item 
    Proposing a methodology to analyze hardware security features for their coverage of the attack space. Additionally, identifying defense features with significant performance overhead (\reftbl{tbl:matrix_defense_mechanism}).

    \item 
    Proposing a new defense feature that allows speculative cache fills with a low-cost invalidation mechanism on the infrequent squash path.

    \item
    Designing a speculation-aware, low-latency, dynamic remapping, randomized cache, inspired by the previous NewCache design\cite{newcache, newcache2016micro}.
    
    \item 
    Designing two Speculative and Timing Attack Resilient (STAR) cache architectures, \cacheinvfa~and \cacheop, defeating all considered cache timing attacks with low performance overhead of 5.6\% and 6.8\% respectively.

\end{itemize}

%% file: 0200_background.tex
\section{Background}
\label{sec_background}

\subsection{Cache Timing Side-channel Attacks}
\label{sec_side_channel}

Cache side-channel attacks observe the cache access time and infer the address used by the victim.

A \bheading{hit-based cache side channel} allows an attacker to infer victim's information by observing cache hits. One common example of hit-based side channels is using the flush-reload technique, e.g., \cite{flushreload2014}, with the following steps:

\begin{enumerate}[leftmargin=*,noitemsep,topsep=0pt]
\item 
(Flush) The attacker first flushes all cache lines of a shared memory region from cache.

\item 
The victim accesses one of the cache lines using a secret-dependent address.

\item 
(Reload) The attacker reloads all cache lines and finds a cache hit at the address just accessed by the victim. The secret can be recovered from this address.
\end{enumerate}

The flush-reload cache side channel is an effective technique which can recover the secret as long as the secret-dependent address is in the cache. While the first step of flushing is through the clflush instruction in the x86 architecture, it can also be achieved in other architectures by causing conflicts to evict related cache lines, leading to a similar variant of the eviction-reload side channel.

A \bheading{miss-based cache side channel} \cite{bernsteinattack, llcfangfei, evictionset2019}, has the victim's secret-dependent access evicting the attacker's cache line(s). The attacker can then recover the secret value by observing where he has a cache miss when accessing the previous lines. The common example of prime-probe attack \cite{primeprobe} leveraging cache contention consists of three steps:

\begin{enumerate}[leftmargin=*,noitemsep,topsep=0pt]
    \item
    (Prime) The attacker fills the cache with his cache lines.
    
    \item 
    The victim accesses a secret-dependent address, which evicts one of the attacker's cache lines.
    
    \item (Probe) 
    The attacker reaccesses his cache lines and finds a cache miss at the cache set that was accessed by the victim. Information is leaked from the evicted address.
\end{enumerate}

Miss-based cache side channel attacks do not require shared memory between the victim and the attacker to succeed, making it a more practical attack than some hit-based side-channel attacks. However, the effectiveness of a miss-based cache side channel can be reduced if there are other memory accesses which introduce unrelated cache conflicts.

\begin{figure}[t]
    \centering
    \includegraphics[width=0.6\linewidth]{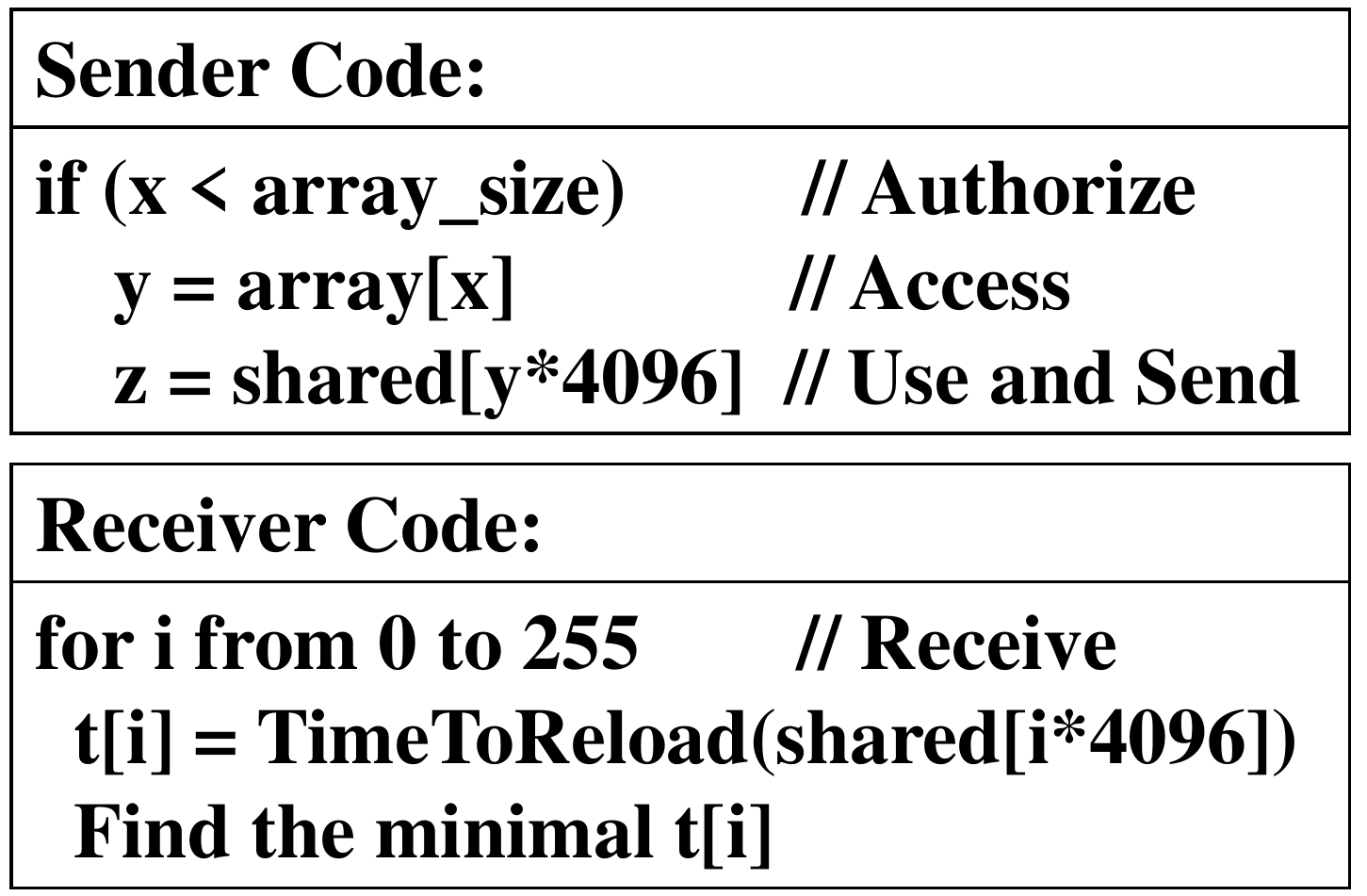}
    \caption{The code of a Spectre-v1 Attack. The mistraining of branch predictor and the flushing of all cache blocks of \textit{shared} happen before the sender code and are not shown.}
    \label{fig:code_spectre_v1}
\end{figure}

\subsection{Speculative Execution Attacks}
\label{sec_bg_spec_attack}

Speculative execution attacks or transient execution attacks are subtle microarchitectural attacks that leak the secret which the software should not access. Research works \cite{AttackSummaryMS, AttackSummary2019Graz, AttackSummary2021Yale, AttackModel2021hpca, AttackSoK2021seed} systematically characterized these speculative execution attacks. There are 2 main operations in a speculative execution attack: illegally accessing a secret, and sending the secret out through a covert channel. \cite{AttackModel2021hpca, AttackSoK2021seed} further broke the attack down to six critical attack steps: \opsetup~of microarchitectural states, \opauth~for bypassed software authorization, \opaccess~for speculative secret access, \opuse~to use the secret, \opsend~and \oprecv~for covert sending and receiving where the cache states can be used.

If the speculation was incorrect, i.e., the access is not authorized, the hardware microarchitecture backs out of the instructions speculatively executed, thus voiding the \opaccess, \opuse~and \opsend~operations. However, since microarchitectural changes, e.g., cache state changes, are supposed to be invisible to software, they are not backed out of. Hence, a cache state change by the \opsend~operation may still be visible, after the incorrect speculation is squashed, thus leaking the secret through the microarchitectural covert channel.

Variants of speculative execution attacks \cite{spectre, spectrev1112, netspectre, spectrersb, ret2spec, spectressb, spectrev3a, lazyfpIntel, meltdown, foreshadow, foreshadowNg, ridl, intelMDS, zombieload, fallout, TAA, VRS, cacheout, intelL1DES, crosstalk, intelSRBDS, lvi, speculativeinterference} exploit different vulnerabilities that allow a transient but illegal access to secret data before this access is verified or authorized. Depending on the vulnerability exploited, the secret can be from special registers, caches, memory or other microarchitectural buffers. All the vulnerabilities can later leverage the cache channel to transmit the secret.

We show an example of Spectre-v1 attack using the flush-reload cache channel in \reffig{fig:code_spectre_v1}. The sender code speculatively reads an out-of-bounds secret into y and executes the covert sending even if x is larger than the array size. The setup step (mistraining for the illegal secret access and flushing for setting up the covert channel) happen before the sender code, and is not shown.

%% file: 0300_threatmodel.tex
\section{Motivation and Threat Model}
\label{sec_threat_model}

\bheading{Cache as a critical leakage channel.}
There are other hardware units that can be used as the microarchitectural channel for both non-speculative and speculative attacks, e.g., the execution port \cite{smotherspectre}, the miss status holding register (MSHR) usage \cite{speculativeinterference} and the branch predictor states \cite{branchscope, nda}. However, the cache state has the following advantages for being a reliable and commonly used channel.

    Distinguishable timing difference. There is a clear difference in timing, e.g., a load instruction takes a few cycles for a L1 cache hit versus hundreds of cycles for a cache miss when the main memory has to be accessed. These make cache timing channels reliable and fast.

    Persistence. The cache stores cache lines that remain after the program finished or terminated. States such as port contention and MSHR usage can be protected by not scheduling concurrent processes. However, the measurement of cache lines can be done by an attack program which runs after the victim program finishes.
    
    High cost to clear. In-core hardware states such as branch predictors can be erased or masked, e.g., with operations like indirect branch prediction barrier \cite{ibpb:amd}, after a program finishes or a context switch happens. However, cache is a hardware unit that has multiple cache levels and a large number of cache lines. These make it a difficult and high-cost operation in hardware to select and clear all cache lines of the victim.
    
    High bandwidth for leakage. Compared to the usage of hardware units which can encode 1 bit of information, the address used to access the cache can encode more bits. \reffig{fig:code_spectre_v1} shows an example of leaking 8 bits. Leaking more bits is possible by using a larger shared array.


\bheading{Threat Model.}
Our threat model considers the attacker that can leak a secret by making a secret-dependent cache state change. The scope includes the cache side-channel attacks where the victim's execution is non-speculative and cache-based speculative execution attacks. Specifically, we protect the addresses of cache lines and cache replacement state. 

In the scope of speculative execution attacks, we consider a load being speculative if not all previous instructions have finished execution without having a fault when the load executes. This covers all forms of malicious speculative execution using different techniques, e.g., predictor mistraining, malicious jump target injection, speculative store-to-load forwarding or delayed exceptions or faults. We assume the strong attacker who can trigger speculative execution even through the victim's own behavior \cite{netspectre}.


We protect the secret from the memory, caches, registers, microarchitectural buffers and other units, i.e., how the address of memory access is computed does not impact the protection. We assume the attacker is able to get timing measurement to detect the presence of his cache lines in various cache levels. Our solution focuses on protecting the level one (L1) data cache, which is close to the processor core and can be easily exploited for cache timing channel. 
The protection of shared last-level cache using techniques like encrypted set indexing \cite{ceaser, evictionset2019, scattercache} is in parallel with our work.

We do not protect against security issues due to software bugs or malicious code injection which give an attacker the permission of legal secret access. We assume the system is able to allocate security domains to software programs automatically or upon programs' request. A security-sensitive program should be assigned a unique DomainID. Security domain ID does not equate to process ID. A given process can have more than one security domain. One security domain can also comprise many processes.

We allow and protect read-only memory regions to be shared between security domains for purposes such as shared software library \cite{sharedlibrary1987auugn}. Sharing writable pages between security domains is dangerous as it allows direct information leakage. 

\begin{table*}[t]
    \centering
    \includegraphics[width=\linewidth]{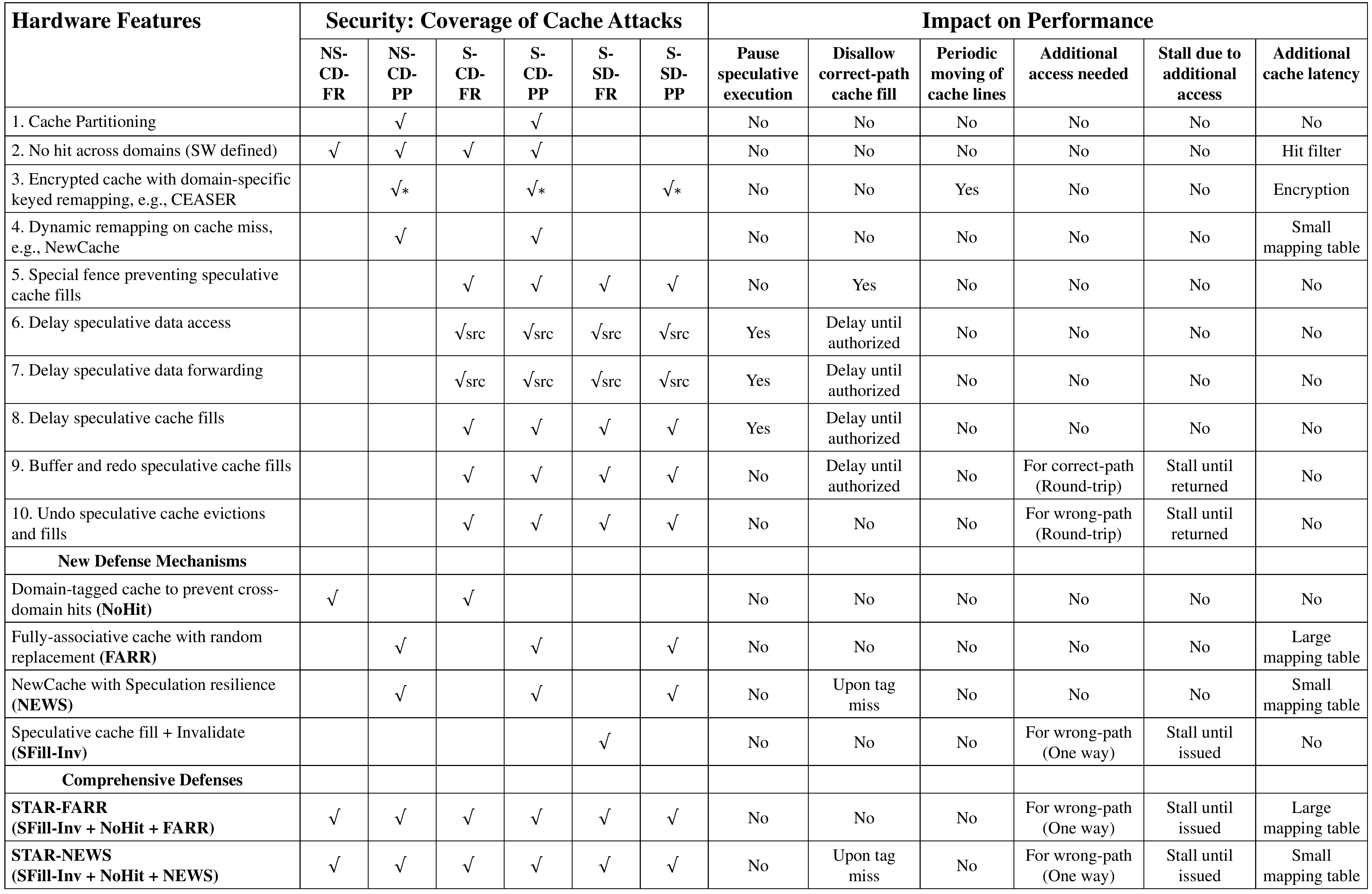}
    \caption{Defense mechanisms against different cache timing attacks. NS is the non-speculative side-channel attack and S is the speculative execution attack. CD stands for a cross-domain attack and SD for a same-domain attack. FR stands for the flush-reload cache channel and PP for the prime-probe cache channel. \checkmark * means the security relies on the design parameter. $\checkmark_{src}$ means the defense protects certain sources of secret, which are usually caches and memory.}
    \label{tbl:matrix_defense_mechanism}
\end{table*}

We do not address physical attacks or circuit-level faults such as Rowhammer-type attacks \cite{rowhammer2014isca, rowhammer2020tcad}.

%% file: 0400_motivation.tex
\section{Methodology}

We propose a methodology to investigate the scope of attacks to be covered (\refsec{sec_types_measurement}), and which mechanisms defeat which attacks (\refsec{sec_security_analysis_past}), and what performance overhead is incurred (\refsec{sec_perf_factor}). Our critical insight from security analysis is that side-channel defenses cannot mitigate speculative execution attacks which do covert sending and receiving in the same security domain. On the other hand, the defenses against speculative execution attacks do not protect the cache state changes by the victim's authorized execution in non-speculative cache side-channel attacks. We also present our observations of the hardware events that cause performance overhead.

\subsection{Similarities and Differences of Attacks} 
\label{sec_types_measurement}

Cache timing attacks include two families of non-speculative (NS) execution attacks and speculative (S) execution attacks. There are two further dimensions: hit-based (exemplified by the Flush-Reload attack, FR) vs miss-based (examplified by the Prime-Probe attack, PP) attacks, and cross-domain (CD) vs same-domain (SD) attacks. This gives rise to the 6 classes of attacks shown by the columns in \reftbl{tbl:matrix_defense_mechanism}.

Both non-speculative and and speculative attack families use either a hit-based or a miss-based channel. The cache states are prepared and measured in the same way (see \refsec{sec_background}). The state change is made by the victim's load either non-speculatively or speculatively.

A critical difference between the non-speculative side-channel attacks and the speculative execution attacks, is that the sender and the receiver can be either in the same domain or different domains. For non-speculative conventional cache side channels, the attacker and the victim are in different security domains. A same-domain attacker with legal permission can access and leak the secret directly instead of leveraging side channels. In speculative execution attacks, the sender and the receiver can be in the same domain or in different domains. The same-domain attack is more dangerous and has been shown to be practical \cite{ret2spec, meltdownRepoIAIK}.

\subsection{Analysis of Past Hardware Defenses}
\label{sec_security_analysis_past}

\reftbl{tbl:matrix_defense_mechanism} shows the security analysis on what cache-based attacks are mitigated by each of the hardware features. Rows 1 to 10 are features implemented by hardware defenses in past work. The columns are the six types of cache timing attacks (see \refsec{sec_types_measurement}).
A checkmark in the table means the hardware feature is able to prevent the corresponding attack.

\bheading{Side-channel defenses.} Secure caches against cache side-channel attacks has two major directions. The first direction is to partition the cache resources between security domains \cite{plrpcache, nomocache, catalyst, mi6, ironhide} (row 1). This method prevents the leakage through secret-dependent cache evictions if the system can separate the victim and the attacker.
Advanced designs, e.g., DAWG \cite{dawg} and Hybcache \cite{hybcache}, implement mechanisms to filter cache hits (row 2) to also prevent the cross-domain hit-based attacks.

The second direction is randomizing the cache change, so the attacker gets no useful information. The keyed remapping in row 3, e.g., CEASER \cite{ceaser} and ScatterCache \cite{scattercache}, uses the domain-specific keys to randomize cache line placement. To prevent eviction set based attacks \cite{llcfangfei, evictionset2019}, the key needs to be changed periodically, which requires existing cache lines to be moved to adapt to the new mapping. The period of rekeying may impact its security against same-domain prime-probe speculative execution attacks (\checkmark * in \reftbl{tbl:matrix_defense_mechanism})  \cite{randcache:analysis2021sp, fixit2021sp, casa}.

Dynamic remapping caches in row 4, e.g., RPCache \cite{plrpcache} and NewCache \cite{newcache, newcache2016micro}, randomly replace cache lines on contention-based evictions, so the attacker cannot get any useful information about the cache lines used by the victim. They can mitigate all miss-based cache side-channels, including the ones based on measuring the time taken by a whole operation (e.g., encryption of a whole block) rather than just the time taken for a single memory access. Since these attacks, e.g., the Evict-Time attack \cite{bernsteinattack}, are much slower, we do not consider them here. NewCache \cite{newcache2016micro} uses the terms Domain ID and P-bit to define security domains for different cache lines. The dynamic remapping effectively does remapping on every cache miss and has not been broken in recent studies of randomized caches \cite{randcache:analysis2021sp, fixit2021sp, casa}. We study if it can be used for a comprehensive solution for both cache side-channel attacks and speculative cache-based attacks.

Previously, dynamic remapping was mainly applied to L1 caches while the keyed remapping was applied to last-level caches (LLC's). We envision the future that both techniques could be used for either cache level, with the emergence of designs for pseudo fully-associative LLC's \cite{mirage} and high-performance hardware lookup tables \cite{capcam}.

We note that side-channel defenses cannot defend against same-domain speculative execution attacks. 

\bheading{Defenses against speculative execution attacks.} Defenses against speculative execution attacks can be categorized as preventing the \opsetup~of microarchitectural states, preventing the secret \opaccess, preventing the \opuse~of secret and preventing the state change by \opsend~operations \cite{AttackModel2021hpca}. 

Protection against malicious \opsetup, e.g., by encrypting or flushing predictor states\cite{samsung:pred:encryp, csf, ibpb:amd}, does not prevent non-speculative cache side-channel attacks and the strong attacker who can trigger speculative execution from the same domain. We do not include it in \reftbl{tbl:matrix_defense_mechanism}.

A defense can delay the speculative execution until it is authorized. Row 6 in \reftbl{tbl:matrix_defense_mechanism} includes mechanisms to delay the speculative secret access, e.g., Context-sensitive Fencing \cite{csf} and secure bounds check \cite{sabc}. Row 7 analyzes defenses delaying the forwarding (use) of secret, e.g., NDA\cite{nda}, SpectreGuard \cite{spectreguard}, ConTExT \cite{context}, SpecShield \cite{specshield} and STT \cite{stt}. These defenses can mitigate multiple covert channels but can only protect specified accesses such as loads of memory or special register reads.

Specific to the cache state, a special fence (row 5) can be inserted to make a speculative load uncacheable to avoid fetching new cache lines \cite{csf}. A defense can delay only speculative loads which have a cache miss (row 8), e.g., Conditional Speculation \cite{condspec}, Efficient Speculation \cite{efficientspec} and DOLMA \cite{dolma}.  Speculative cache lines can also be put in a special buffer (row 9) and made visible once it is authorized, e.g., InvisiSpec \cite{invisispec}, SafeSpec \cite{safespec} and MuonTrap \cite{muontrap}.

To improve the performance with the insight that most of the instructions in benign programs are not squashed, a defense can also allow speculative cache fills and restore the cache state if the execution is squashed (row 10). CleanupSpec \cite{cleanupspec}, for each squahsed loads, removes the speculatively fetched cache line and refetches the evicted cache line. However, CleanupSpec needs to wait until the restoration is finished. The wait time is found to be secret-dependent and measureable in the unXpec attack \cite{unxpec}.

These defenses can mitigate cache-based speculative execution attacks either from the same domain or a different domain. However, these defenses against speculative execution attacks will not defeat non-speculative side-channel attacks as the victim's execution will not be squashed.

\bheading{Takeaways.}
We show that none of the existing cache defenses can cover both the same-domain speculative execution attacks and the non-speculative side-channel attacks. Hence, our goal is to design a secure cache that covers all of these attacks with minimal performance impact and hardware complexity.

\begin{figure}[t]
    \centering
    \includegraphics[width=\linewidth]{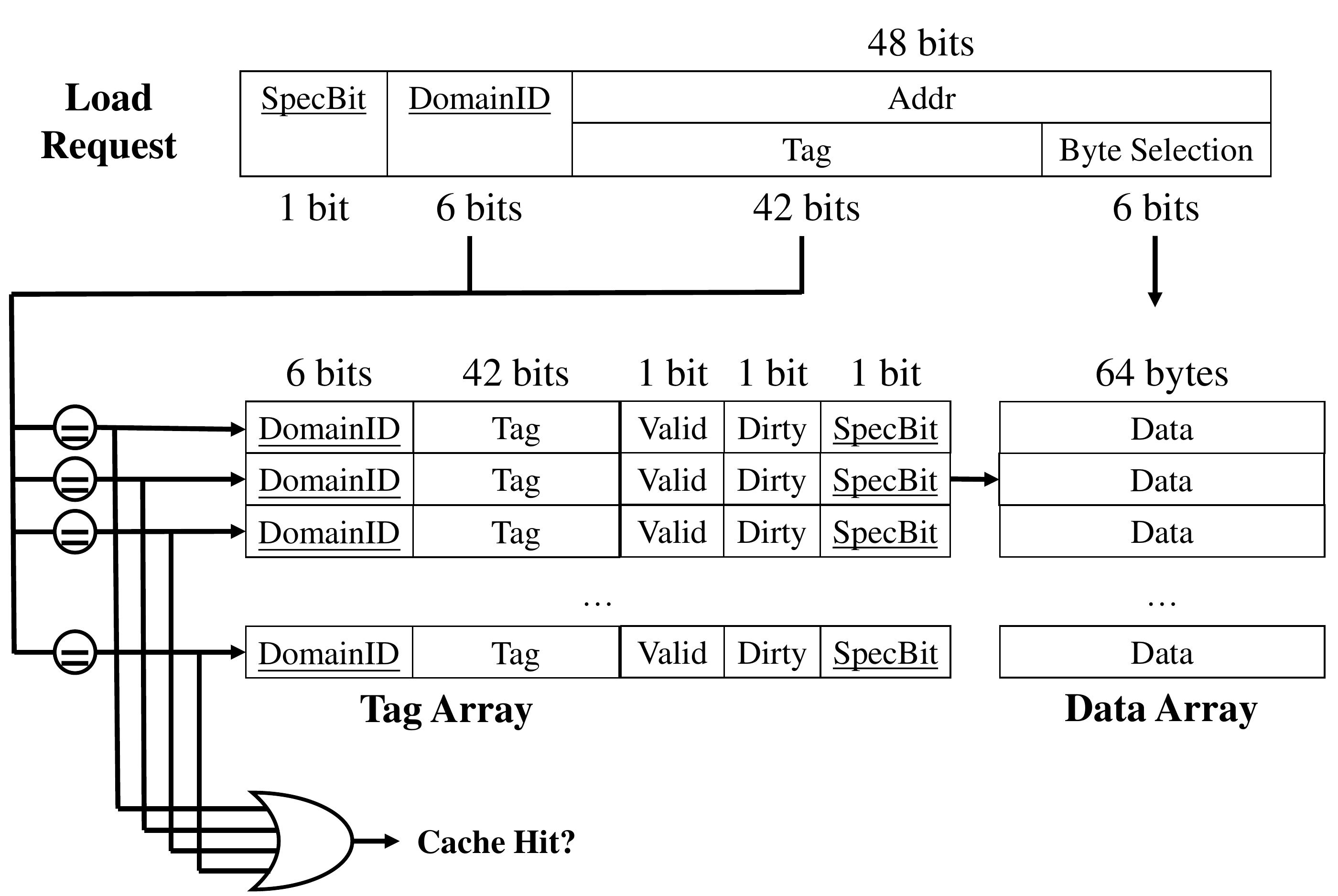}
    \caption{Architecture of \cacheinvfa. DomainID and SpecBit are attached to memory requests and cache lines to indicate the owner and the speculation status.}
    \label{fig:cache_arch_fa}
\end{figure}

\subsection{Factors of Performance Overhead}
\label{sec_perf_factor}

We identify six types of hardware events which lead to performance degradation by the defenses in \reftbl{tbl:matrix_defense_mechanism}. A good defense should try to reduce the possibility of all these events happening while maintaining the coverage of attacks.

\bheading{Delay correct-path execution.}
Defenses on rows 6 to 8 delay the execution and related cache fills of speculative instructions. This can cause severe performance overhead as the correct-path execution is more frequent.

\bheading{Disallow correct-path cache fill.}
Cache fills are critical for performance.
While permitting speculative data accesses, fill-prevention fences on row 5 may disallow a large portion of cache fills if implemented with the strict threat model.

\bheading{Periodic moving of cache lines.}
Periodic moving of existing cache lines is required as the keyed remapping defense (row 2) changes the mapping of address to cache line. It is a heavy-weight hardware operation to examine existing cache lines for their ownership and perform manipulation. 

\bheading{Additional access needed.}
Defenses using speculative buffers (row 9) need a second access to redo accesses for more frequent correct-path execution. It is a round-trip operation in cache to access the address and refetch it into the cache. Fill-and-undo defenses (row 10) need to both invalidate the speculatively accessed address and refill the evicted address if the speculative execution is squashed (wrong-path).

Undo-based defenses can be intrinsically better in terms of performance overhead since they do extra work on the less frequent, wrong speculation path. However, CleanupSpec has to bring back the evicted cache line to the L1 cache which is complicated and timing consuming.

\bheading{Stall due to additional access.}
Defenses using speculative buffers (row 9) need to stall the commit of speculative instructions as the validity of speculatively accessed data needs to be verified. Fill-and-restore defenses (row 10) also need to stall the execution of later memory operations before the restoration is finished.

\bheading{Additional cache latency.} Hardware units required by defenses may add to the cache access latency time, e.g., encryption for randomization and large fully-associative caches.

%% file: 0500_architecture.tex
\begin{figure}[t]
    \centering
    \includegraphics[width=\linewidth]{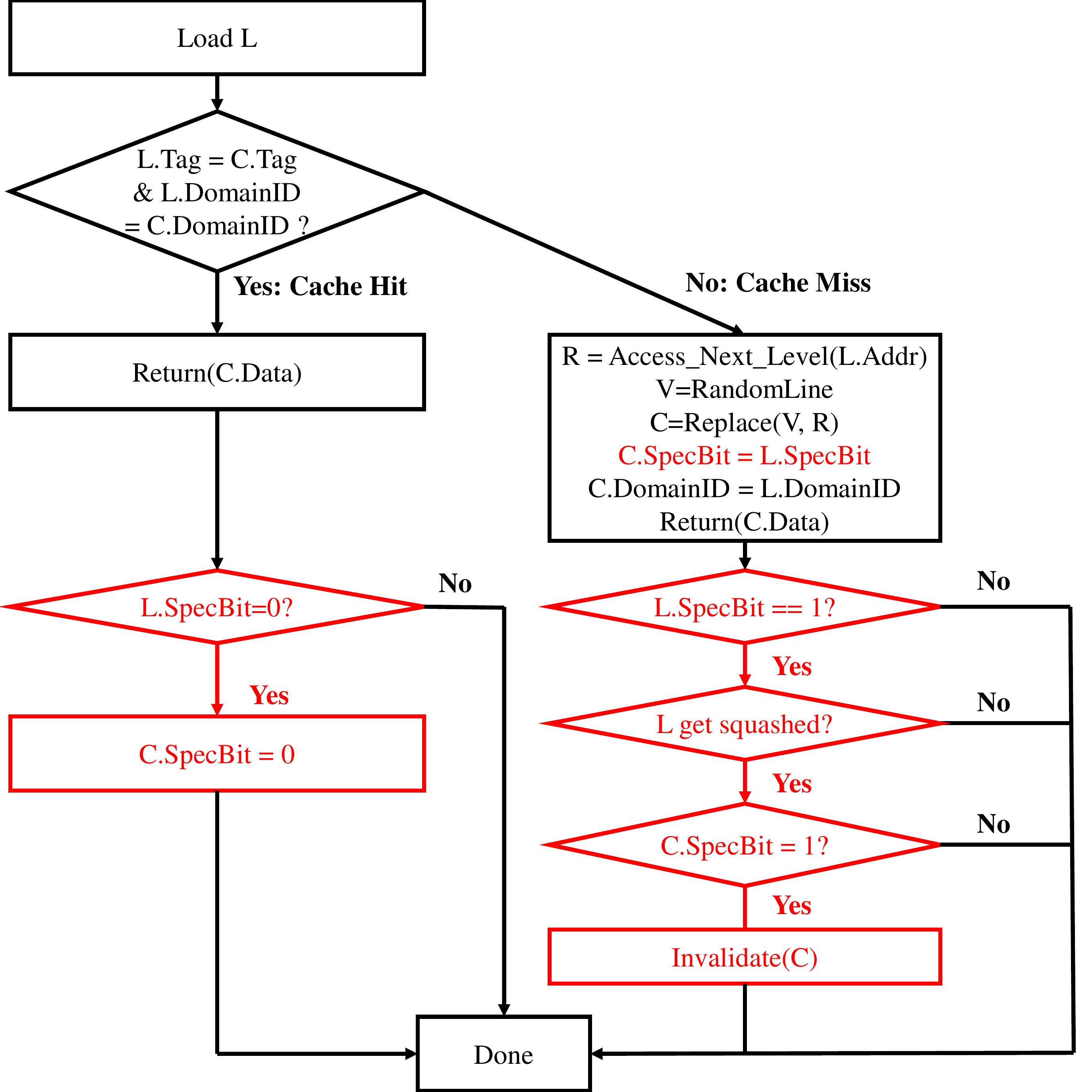}
    \caption{Load handling in \cacheinvfa. Modifications to handle speculative execution attacks are shown in red.}
    \label{fig:access_policy_sqinv_fa}
\end{figure}

\begin{figure*}[t]
    \centering
    \includegraphics[width=0.95\linewidth]{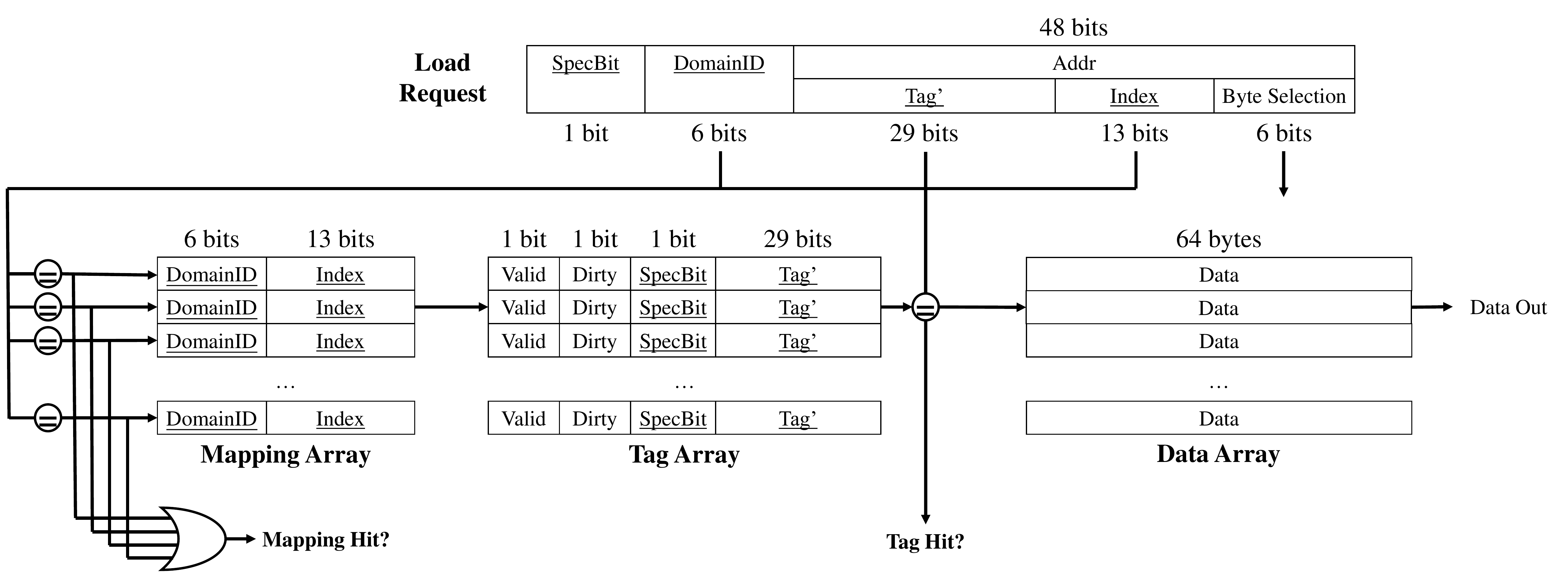}
    \caption{Architecture of NewCache-inspired \cacheabbrev~(\cacheop). New and modified fields of load requests and cache lines are underlined. For a 32kB L1 cache with 64-byte cache line size, there are with $2^9$ = 512 cache lines so an index width of 9 bits is required. By making the index field wider by adding k bits (k = 4 in this example), it is like mapping to a logical cache that is $2^k$ times larger. If a 48-bit address is used, the modified \textit{Index} field has 9 + 4 = 13 bits while the \textit{Tag'} field is 42 - 13 = 29 bits.}
    \label{fig:cache_arch}
\end{figure*}

\section{\cacheabbrev~Architecture}
\label{sec_arch}

\subsection{Overview}

\cachename~is a comprehensive defense against both speculative execution and side-channel attacks. To cover the non-speculative cache side-channel attacks in \reftbl{tbl:matrix_defense_mechanism}, we suggest using some form of randomized cache for contention-based PP attacks and a security domain field to disallow cache hits across domains for FR attacks. These cover all the attack columns in \reftbl{tbl:matrix_defense_mechanism} for both non-speculative and speculative execution attacks except for the same domain speculative flush-reload attacks. We propose a low-cost and high-performance solution to this (\sqinv).

We introduce the following four new defense features and two variants of \cacheabbrev~protecting the L1 cache. These build upon fully-associative cache with random replacement (\cacheinvfa~in \refsec{sec_cacheinvfa}) and a new speculation-aware randomized cache (\cacheop~in \refsec{sec_cacheop}). They are shown to defeat all columns of attack types and compared with existing defenses in \reftbl{tbl:matrix_defense_mechanism}


\bheading{Domain-tagged cache to prevent cross-domain hits (\hitfilter).} We attach DomainID's to cache lines as well as internal buffers such as miss status holding registers and write-back buffers. Memory accesses from a different domain cannot get a hit on a cache line with a different DomainID, which defeats different-domain flush-reload attacks. While the existing defense of software-defined hit filter \cite{dawg} requires the system software to be aware of hardware configurations and manually allocate resources for each domain, the domain-tagged cache only requires the system software to assign DomainID's and the cache allocation is done automatically by hardware.


\bheading{Fully-associative cache with a random replacement policy(\farr).} We show that the \farr~, as a basic cache architecture, provides security against miss-based cache attacks. This is because cache lines of the attacker have the equal chance to be replaced no matter what address is accessed. \farr~also prevents the leakage through the cache replacement state.

\bheading{Speculation-aware NewCache (\sdr).} NewCache \cite{newcache, newcache2016micro} has the same security profile as \farr. FARR is conceptually simpler but may increase access latency. The original NewCache has been shown to have the same access time as a same-size set-associative cache \cite{newcache2015asscc}. Inspired by NewCache, we propose a speculative-aware cache with dynamic remapping. We show adjustments for security in \refsec{sec_cacheop}.

\bheading{Speculative cache fill and invalidate on squash (\sqinv).} 
We propose \sqinv, a new mechanism to invalidate speculatively fetched cache lines upon a squash. This is a new feature that can improve the security and also reduce performance overhead compared to the previous undo-type defense such as CleanupSpec\cite{cleanupspec}. For security, \sqinv~does not require the processor to wait until completion, eliminating the root cause of the unXpec \cite{unxpec} attack on CleanupSpec due to measureable restoration time. For performance, it does only invalidation on squash and does not require bringing back the evicted cache line. The details of \sqinv~is described in \refsec{sec_sqinv}.

\begin{figure*}[t]
    \centering
    \includegraphics[width=0.92\linewidth]{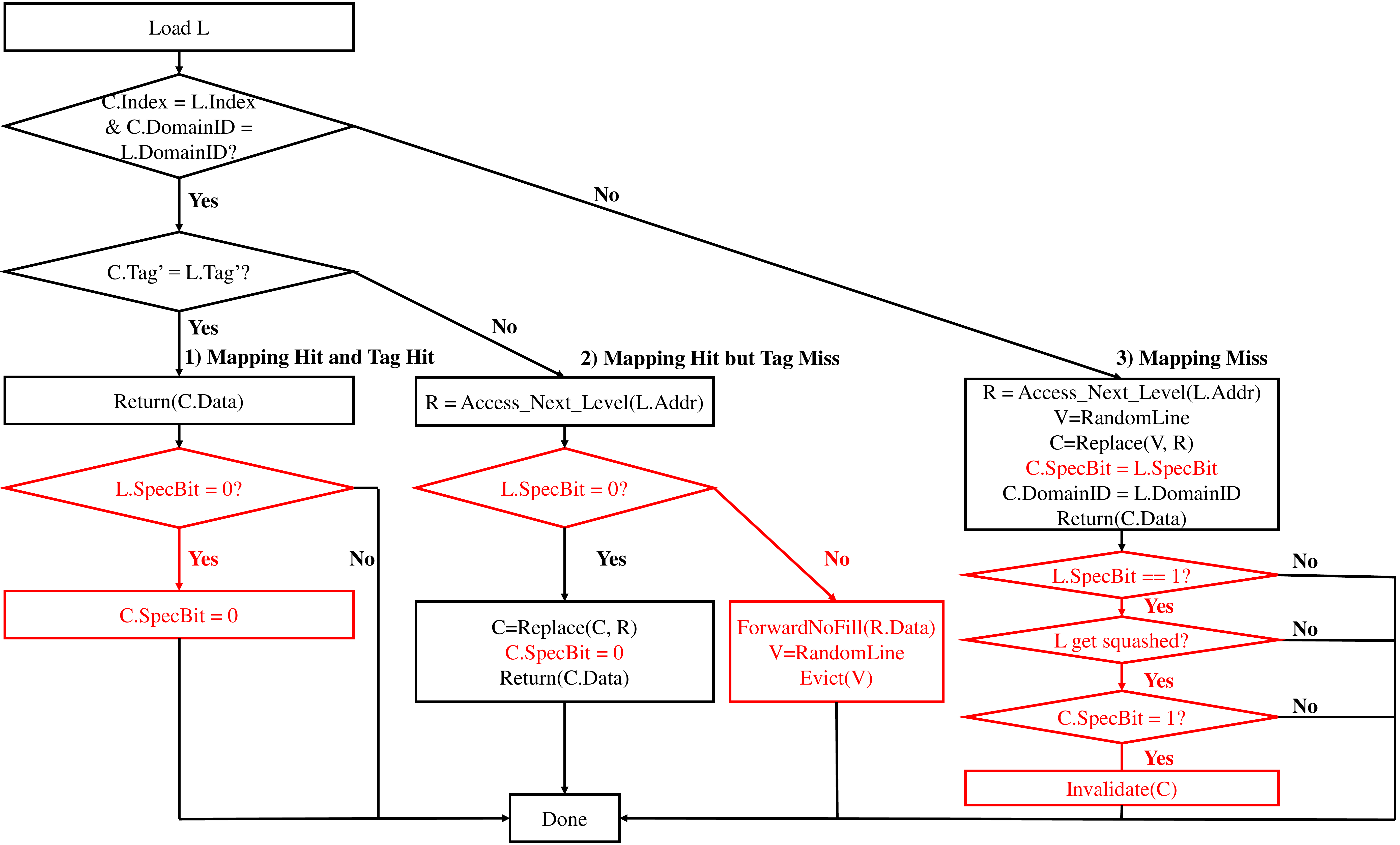}
    \caption{Load handling for \cacheop~. Modifications to handle speculative execution attacks are shown in red.}
    \label{fig:access_policy}
\end{figure*}

\subsection{\cacheinvfa}
\label{sec_cacheinvfa}

\cacheinvfa~implements speculative cache fill + invalidation (\sqinv), preventing cross-domain hit (\hitfilter) and a fully-associative cache with random replacement(\farr). \cacheinvfa~defeats all the attack columns in \reftbl{tbl:matrix_defense_mechanism}.

The hardware modifications of \cacheinvfa~are shown in \reffig{fig:cache_arch_fa}. A load request records in \textit{SpecBit} whether it is speculatively executed and considered insecure. Store operations are always issued when it can no longer be squashed so the SpecBit of stores will always be 0. \textit{DomainID} field in the request and in the cache line denotes the security domain of the memory operation and the cache line respectively.

\reffig{fig:access_policy_sqinv_fa} shows the load handling in \cacheinvfa. A prefix of \textit{L.} refers to the load request, while a prefix of \textit{C.} refers to a cache line. \cacheinvfa~checks the \textit{Tag} and the \textit{DomainID} to decide whether there is a cache hit. A non-speculative request can clear the \textit{SpecBit} of the cache line at which it has a hit. A missing request can randomly replace an old cache line even if it is speculative. If the speculative load is squashed later, a \sqinv~request is sent to the cache to invalidate the address.

\subsection{\cacheop}
\label{sec_cacheop}

Using fully-associative caches in \cacheinvfa~can increase the latency of an access, the circuit size and the power consumption. \cacheop~is the optimized architecture to reduces access time latency and power consumption by adopting the enhanced NewCache with speculation resilience (\sdr) instead of the fully-associative cache.

\cacheop~architecture is shown in \reffig{fig:cache_arch}. The new or modified fields of load requests and cache lines are shown as underlined items. In a NewCache-like archiecture, the \textit{Tag} field in a \farr~(\reffig{fig:cache_arch_fa}) is divided into a shorter \textit{Index} field and a \textit{Tag'} field. The power needed to compare with every long \textit{Tag} in \farr~is reduced because the parallel comparisons are only with the shorter \textit{Index} field. When both \textit{DomainID} and \textit{Index} fields match on an cache access, the comparison is done with the \textit{Tag'} of that cache line.

\cacheop~allows allocating more bits in the address as index bits to represent ``a bigger logical cache (than the actual physical cache size)'', which is later shown to have further performance benefits. In \reffig{fig:cache_arch}, we show an example with 4 extra bits allocated to the \textit{Index} field. The mapping entry of a cache line consists of \textit{DomainID} and \textit{Index}, giving a dynamic address-to-cache remapping for that domain.

\reffig{fig:access_policy} shows the load handling in \cacheop. For memory accesses, \cacheop~first checks the mapping entry array to see if there is a cache line with the matching \textit{DomainID} and Index bits. 

A mapping miss happens if there is no cache line that has the same mapping entry bits as the request. The tag array is only checked if the mapping entry array has a \textbf{Mapping Hit}. A cache hit is when the cache has both a mapping and a tag hit. If there is a mapping hit but the \textit{Tag'} fields do not match, a tag miss happens. We show that the handling of a load request in different paths will not leak information and cannot be used to measure existing cache states.

\bheading{Mapping hit and Tag Hit.}
For a cache hit, \cacheop~returns the data as a normal hit in conventional caches. As a side effect, if the cache line was fetched by a previous speculative memory operation (with the \textit{SpecBit} set) and is accessed by a non-speculative operation, its \textit{SpecBit} is cleared.

Introducing the comparison of \textit{DomainID} when looking for a mapping hit means cache lines of shared memory region are not shared. In other words, a program from another security domain cannot get a hit at the current security domain's cache line even if the address bits match. This \hitfilter~feature prevents the hit-based side-channel attacks since the attacker cannot observe a hit at the address used by the victim. 


\bheading{Mapping Hit but Tag Miss.}
For a mapping hit at a cache line C but a miss for \textit{Tag'}, it needs to fetch the cache line from the next level of memory. If the load is a non-speculative load with its \textit{SpecBit} cleared, when the requested cache line R is returned, it is allowed to replace C. At the same time, \cacheop~will clear the \textit{SpecBit} of R and return the data.

However, for a speculative load, replacing the cache line C can cause leakage of the address if C was placed for a same-domain speculative execution attack. Evicting C enables the attacker to infer the \textit{Index} bits of the load address. \cacheop~disallows filling the cache with R and forwards the data, which causes no leakage. 
In addition, \cacheop~needs to evict a random cache line as keeping all old lines can cause an observation different from the handling of a mapping miss which evicts one of the lines. A strong attacker can try to construct eviction sets \cite{llcfangfei, evictionset2019} from this difference. Handling speculative loads and non-speculative loads differently allows \cacheop~to defeat the same-domain miss-based attack (S-SD-PP in \reftbl{tbl:matrix_defense_mechanism}) which the original NewCache cannot prevent. This also shows that \textit{designing a comprehensive design is not just combining different security features, and adjustment should be made to avoid introducing new attacks}.

Although \cacheop~protects the security, evicting a cache line but not filling the cache can lead to performance overhead. \cacheop~can reduce the frequency of not filling by having more bits in the index, which makes it less likely for \cacheop~to have a mapping hit and tag miss. It also allows the flexibility of performance tuning (see \refsec{sec_perf_eval}) while maintaining the security and the physical cache size.

\bheading{Mapping Miss.}
For a mapping miss, the next level of memory is also accessed. When the requested cache line R is returned, a random cache line V is replaced by R. The \textit{SpecBit} and the \textit{DomainID} of R are then set to be the same as the load request. If the load is speculative and gets squashed later, the processor will send a \textit{\sqinv} signal to the cache to invalidate the speculatively fetched cache line R. If R has been accessed by some non-speculative memory operations and has its \textit{SpecBit} cleared before the squash, R can be preserved.

In a miss-based side-channel attack, the victim will always have a mapping miss and replace the attacker's cache lines as their \textit{DomainID}'s do not match. The randomized replacement in \cacheop~prevents the leakage as the victim's line can evict any of the attacker's lines with the same probability.

Similar protection is offered against speculative execution attacks. 
A miss-based different-domain attacker can only see a random eviction and learn nothing about the address of the speculative load L. 
For same-domain attacks, if the sender's speculative load has a mapping miss, it still randomly replaces a line, leaking no information to the attacker.

\bheading{Take-aways.} \cacheop~achieves a low-latency speculation-aware randomized cache. While inspired by NewCache, it further solves the subtle but critical same-domain prime-probe attack exploiting the mapping hit case.

%% file: 0600_impl.tex
\subsection{Implementation of \sqinv}
\label{sec_sqinv}


We optimize the performance of our new ``Speculative cache fill + Invalidate
(\sqinv)'' feature to reduce the pipeline stalls and round-trip memory traffic. We describe the modifications to the memory request, memory response, the squash procedure of loads and the handling of \sqinv~in cache.

\bheading{Memory Request and Response.} 
Each memory response is extended with an extra \textit{SourceLevel} field. \textit{SourceLevel} records the cache level where a memory access finds the requested address. In a two-level cache system, \textit{SourceLevel} is set to 1 for a L1 cache hit. Similarly, \textit{SourceLevel} is set to 2 for L2 cache hit, and 3 if the the data is from memory.

\bheading{Load Squash Procedure} Upon a pipeline squash, squashed loads are required to send a \sqinv~request to invalidate the cache line it fetched. The \sqinv~request is skipped if the speculative load had an L1 cache hit and the returned \textit{SourceLevel} was 1. \textit{SourceLevel} is attached to the \sqinv~request and used for handling. Compared to defenses which need to stall the processor until the response of the second access is returned (\refsec{sec_perf_factor}), \sqinv~resumes the execution as soon as all the \sqinv~requests are sent. The subtle unXpec \cite{unxpec} attack trying to observe the restoration time will not work as the processor only sends the \sqinv~request without knowing when it will be completed.

\bheading{Handling of \sqinv.}
We implement a low-cost \sqinv~operation which does not fetch a new cache line or generate a response. This makes the \sqinv~operation a one-way request, reducing half of traffics in the cache system compared to a round-trip memory request. Our implementation is based on an inclusive and write-back cache system.  
When a cache receives a \sqinv~request, it 
looks up the address of \sqinv, which have three possible results.

If the address is found and the \textit{SpecBit} of the cache line is 0, this means the cache line has been accessed by non-speculative memory accesses. This cache line is considered safe and the \sqinv~request can be safely dropped. 

If the address is found and the \textit{SpecBit} of the cache line is 1, the cache line is invalidated. The cache then checks the \textit{SourceLevel} of \sqinv~and sends the request to the next level if \textit{SourceLevel} is larger than the current cache level.

If the address is not found, this could happen if the speculatively fetched cache line gets replaced or evicted before the squash. The request will propagate to 
the next cache level if \textit{SourceLevel} is larger than the current cache level.



\subsection{Performance Impact}

Considering the performance factors in \refsec{sec_perf_factor}, \cacheinvfa~and \cacheop~neither affect frequent correct-path execution nor require periodic moving of cache lines. They only perform extra work when a squash happens (infrequent wrong path) to send a light-weight \sqinv~request without the need to bring back the evicted cache line. Pipeline stall time is also reduced by not waiting for \sqinv's completion.

\begin{table}[t]
    \centering
    \includegraphics[width=\linewidth]{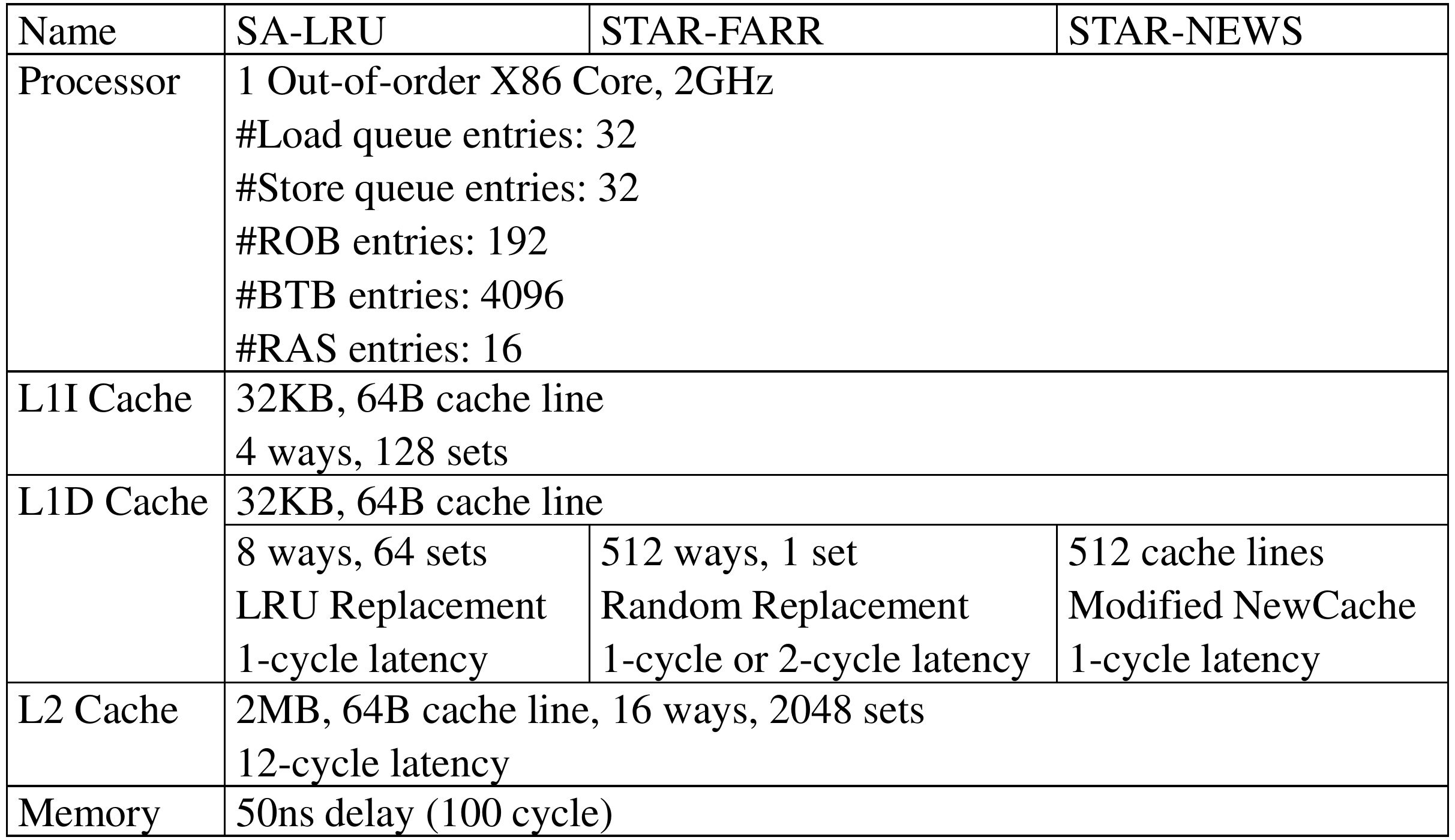}
    \caption{Hardware Configurations of GEM5 Simulator.}
    \label{fig:hw_config}
\end{table}

\begin{figure*}[t]
    \centering
    \includegraphics[width=\linewidth]{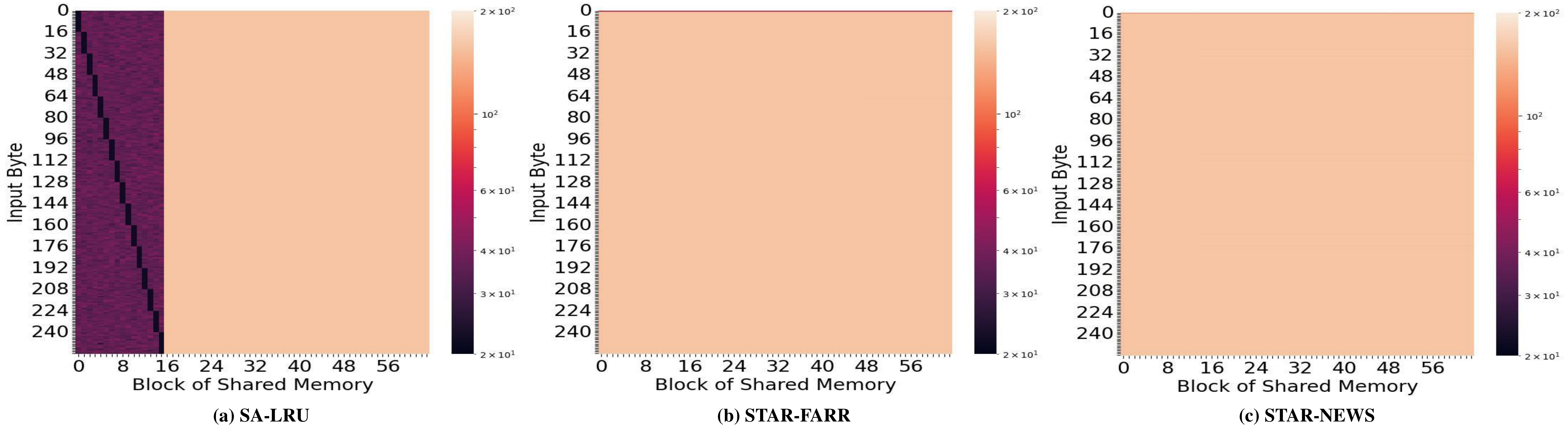}
    \caption{Flush-reload side-channel attack on AES. Lighter is longer cache access time. The key byte is 0, and the dark diagonal shows its XOR result with the input byte which leads to shorter execution time.}
    \label{fig:eval_aes_fr}
\end{figure*}

\begin{figure*}[t]
    \centering
    \includegraphics[width=\linewidth]{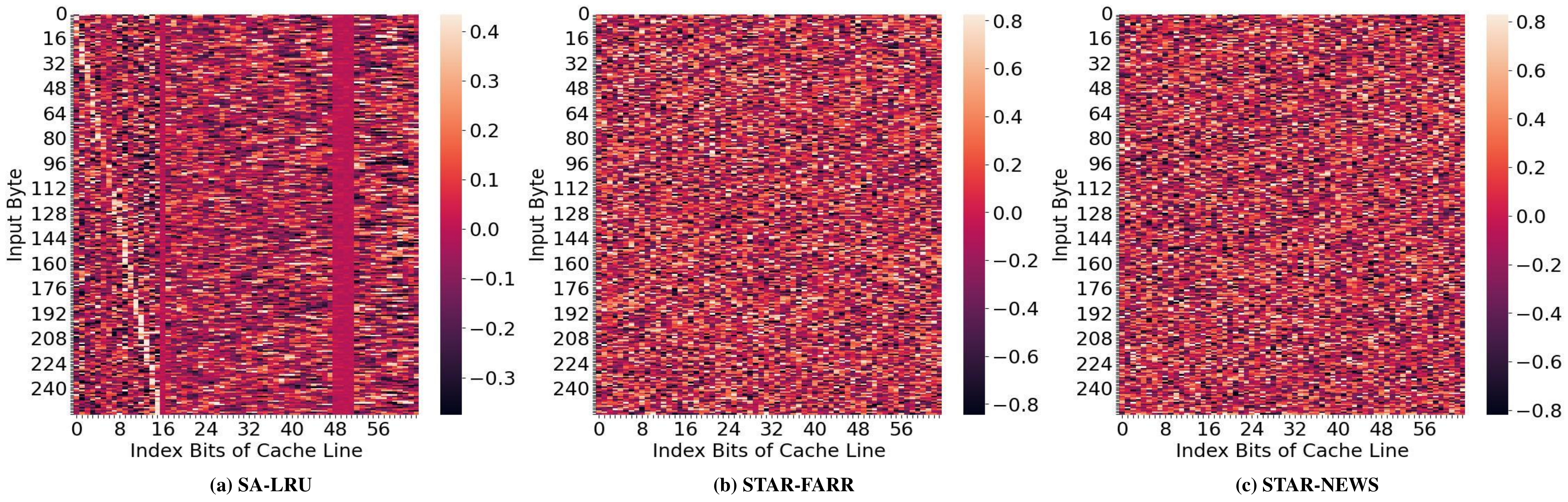}
    \caption{Prime-probe side-channel attack on AES. Lighter is longer cache access time. The key byte is 0, and the light diagonal shows its XOR result with the input byte which leads to longer execution time.}
    \label{fig:eval_aes_pp}
\end{figure*}

%% file: 0700_security_eval.tex
\section{Evaluation}
\label{sec_eval}



We evaluate the security and the performance of hardware architectures whose design parameters are shown in \reffig{fig:hw_config}. We use similar size and latency parameters as InvisiSpec \cite{invisispec} and CleanupSpec \cite{cleanupspec} to get comparable results. The architectures are implemented in the cycle-accurate GEM5 simulator \cite{gem5}. We evaluate the set-associative cache with least-recently-used replacement (SA-LRU) as the baseline. 

\cacheinvfa~and \cacheop~are our proposed defenses which defeat all attacks. As the access latency of a fully-associative cache cannot always be as small as the set-associative cache, we run experiments of STAR-FARR with both a fast design with 1-cycle latency and a slow design with 2-cycle latency. The access latency of NewCache-type L1 cache in \cacheop~can be as low as the set-associative cache in real circuits\cite{newcache2015asscc}. The L2 cache is a set-associative design representing the last-level cache (LLC). The latest secure LLCs using randomization \cite{ceaser, scattercache} are also set-associative so the configuration can also model the timing with LLC protection before a remapping of cache sets is performed.


The security evaluation uses the following 4 representative attacks covering side-channel attacks and speculative execution attacks: (1) A flush-reload side-channel attack on the Advanced Encryption Standard (AES) algorithm. (2) A prime-probe side-channel attack on AES. (3) A Spectre-v1 speculative execution attack leveraging the flush-reload cache covert channel. (4) A Spectre-v1 speculative execution attack leveraging the prime-probe cache covert channel. For the two spectre-v1 attacks, we use a same-domain attacker which is more dangerous than a cross-domain attack.

\begin{figure*}[t]
    \centering
    \includegraphics[width=0.97\linewidth]{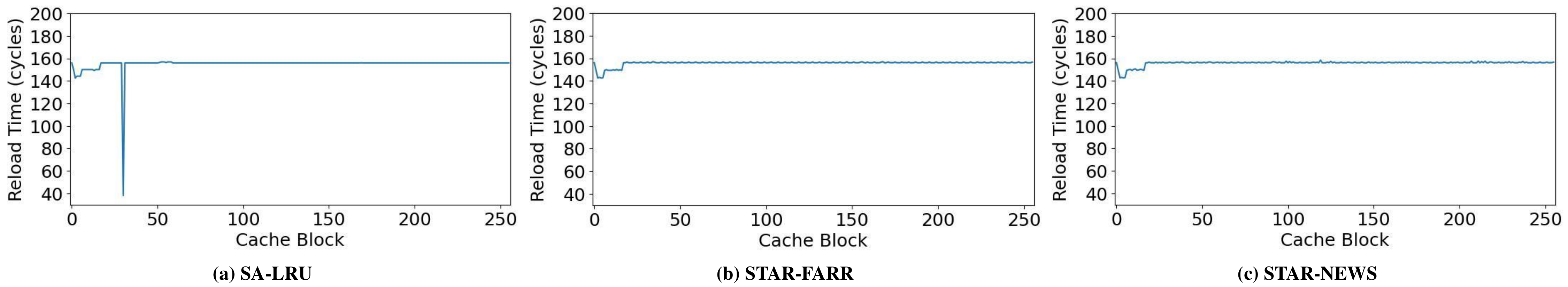}
    \caption{Flush-reload Spectre v1 attack. The secret is 30.}
    \label{fig:eval_spectre_fr}
\end{figure*}

\begin{figure*}[t]
    \centering
    \includegraphics[width=0.97\linewidth]{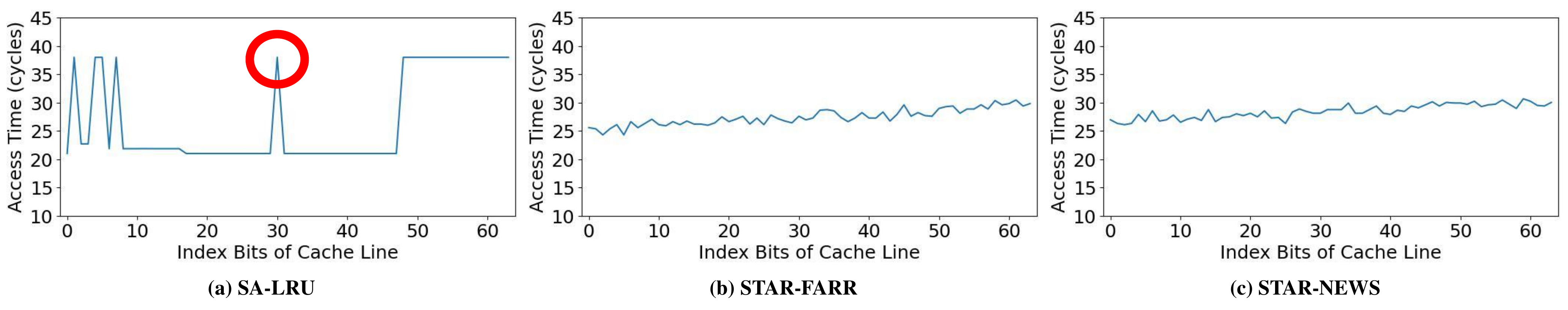}
    \caption{Prime-probe Spectre v1 attack. The secret is 30.}
    \label{fig:eval_spectre_pp}
\end{figure*}

\begin{figure*}[t]
    \centering
    \includegraphics[width=0.95\linewidth]{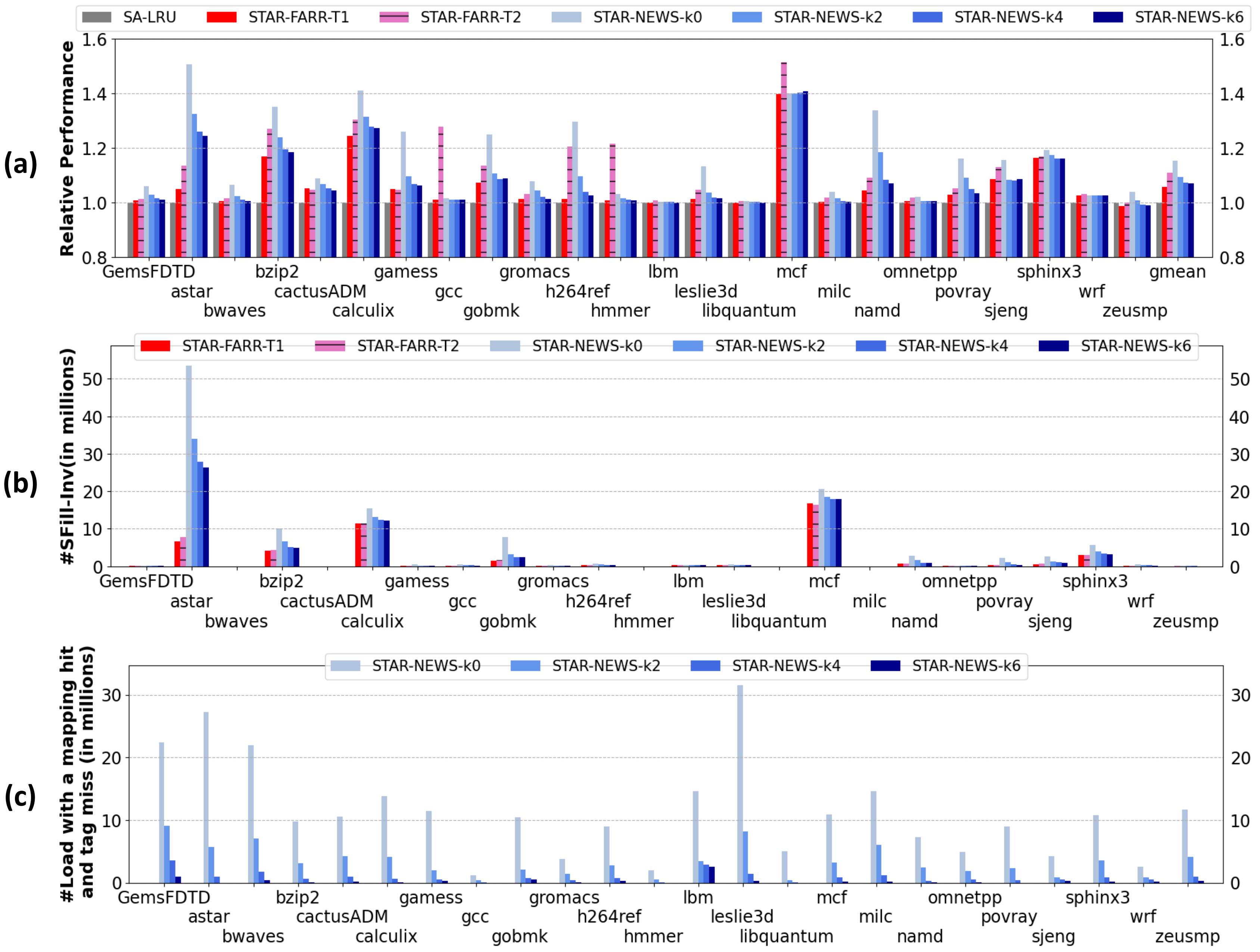}
    \caption{Statistics for performance evaluation: (a) Overall performance: relative execution time of tested architectures running different benchmarks. The last set is the geometric means of relative performance. (b) The number of \sqinv~requests sent to the cache system. (c) The number of speculative loads which have a mapping hit but tag miss in \cacheop~systems.}
    \label{fig:eval_perf}
	\vspace{-5pt}
\end{figure*}

\subsection{Security: Side-channel Attack}

We evaluate the side-channel attacks on the first round of the AES-128 encryption process. 
Optimized AES implementations \cite{aes2002design} compute its operations, \textit{SubBytes}, \textit{ShiftRows} and \textit{MixColumns}, using pre-computed lookup tables. As the read addresses to lookup tables are decided by the encryption key, a side-channel attacker can recover bits in the secret key by analyzing memory access patterns to the cache.

In the first round, the encryption algorithm reads 4 lookup tables, each of which have 256 4-byte entries. We denote the 16 bytes of input data as $D_i$ and the 16 key bytes as $K_i$ with i from 1 to 16. The accesses to lookup tables $T_1$ to $T_4$ are: 

$T_1[D_1 \oplus K_1], T_1[D_5 \oplus K_5], T_1[D_9 \oplus K_9], T_1[D_{13} \oplus K_{13}]$

$T_2[D_2 \oplus K_2], T_2[D_6 \oplus K_6], T_2[D_{10} \oplus K_{10}], T_2[D_{14} \oplus K_{14}]$

$T_3[D_3 \oplus K_3], T_3[D_7 \oplus K_7], T_3[D_{11} \oplus K_{11}], T_3[D_{15} \oplus K_{15}]$

$T_4[D_4 \oplus K_4], T_4[D_8 \oplus K_8], T_4[D_{12} \oplus K_{12}], T_4[D_{16} \oplus K_{16}]$

Each access to the lookup table brings a cache block that contains this address into the cache. We verify the security against a strong attacker who knows or controls the input data to the AES algorithm and tries to recover the key by executing memory accesses and observing timing differences. The side-channel attack on AES is repeated by $2^{15}$ times with random input data and the receiver's timing measurements are averaged as the final results.

\bheading{Flush-reload Side-channel Attack.} A flush-reload attack can happen when the AES table is in a memory region shared by the attacker and the victim. To recover one key byte, the attacker flushes lookup table entries and waits for the victim to execute. As key bytes $K_1, K_5, K_9$ and $K_{13}$ are used to access $T_1$ in the first round, the attacker can infer the value of these bytes by reloading all $T_1$ entries and observing caches hits. Similarly, the attacker can flush and reload entries in $T_2, T_3$ and $T_4$ and recover other key bytes.

\reffig{fig:eval_aes_fr} (a) shows a successful flush-reload side-channel attack on AES. The experiment represents a shared 4kB memory region (64 cache blocks as the x axis) whose first part saves a 1kB AES lookup table. The y axis is the value of the input byte $D_1$. The light region is never accessed in the AES encryption so reloading these blocks takes the longest time. The first 16 cache blocks contain the AES table entries and have shorter average access latency. There is a dark diagonal in the figure, meaning that a cache block has even shorter reload time when the input byte is of certain values. When the value of input byte is from 0 to 15 (0x00 to 0x0f), the first block of AES table $T_1$ (offset: 0x00) is accessed. As the access to $T_1$ in the first round is $T_1[D_1 \oplus K_1]$, $K_1$ is partially leaked to be 0x0u (u is the unknown 4 bits as the 16 lookup table entries in the same cache line cannot be distinguished ).

In \cacheinvfa~and \cacheop, the victim AES encryption and the flush-reload attacker have different DomainID's so any cache lines used by the victim will not cause a cache hit in the attacker's reloading (see \reffig{fig:eval_aes_fr} (b) and (c)).

\bheading{Prime-probe Side-channel Attack.} A prime-probe attack does not require shared AES lookup tables. The attacker fills the cache with the cache lines of his array. The lookup table access, e.g., $T_1[D_1 \oplus K_1]$, will replace cache lines in one cache set determined by $D_1$ and $K_1$, which causes a longer latency when the attacker probes this set later. 

\reffig{fig:eval_aes_pp} (a) shows a successful prime-probe side-channel attack on a AES key byte which is 0. For each cache set (x axis), the heatmap shows whether certain input values lead to longer access time than the average. The light line shows The attacker's cache lines in the first 16 cache sets are shown to conflict with AES table entries, leading to larger access time. Similar to the previous flush-reload attack, the key byte can be recovered as 0x0u (u is the unknown bits).

The fully-associative cache with random replacement in \cacheinvfa~defeats the prime-probe attack on AES (see \reffig{fig:eval_aes_pp} (b)). The dynamic remapping cache in \cacheop~also guarantees a random replacement as a DomainID mismatch always leads to the path of \textit{Mapping Miss} in \reffig{fig:access_policy}, defeating the attack in \reffig{fig:eval_aes_pp} (c).

\subsection{Security: Speculative Execution Attack}

The Spectre v1 attack \cite{spectre} exploits misprediction for conditional branches. The sender transiently bypasses the bounds checking which prevents illegal out-of-bounds accesses. A cache-based Spectre v1 sender transiently reads an out-of-bounds secret and uses the secret to generate a memory access that will fetch or evict a cache line observable by the receiver.

\bheading{Flush-reload Speculative Execution Attack.} The gadget shown in \reffig{fig:code_spectre_v1} leaks a 8-bit secret through the flush-reload covert channel. The sender code in the conditional branch is executed due to branch misprediction. The secret at array[x] is accessed using an illegal offset x 
and used to access a cache line in the \textit{shared} array. The receiver can infer the secret value by looking for the $i$ which gives the shortest reload time.

\reffig{fig:eval_spectre_fr} shows the results of running the flush-reload Spectre v1 attack when the secret is 30. For the SA-LRU baseline (blue), the corresponding short access time at the 30th block, leaking the secret which is 30. \cacheinvfa~(orange) and \cacheop~(green) have no observable low reloading latency as the speculatively fetched cache line is cleared by the \sqinv~feature.

\bheading{Prime-probe Speculative Execution Attack.} In a prime-probe Spectre v1 attack, the attacker uses its array to fill the cache. The Spectre v1 sender gadget will access the secret and evict a cache line when using the secret as the address. The access latency to different cache sets are later accumulated by the receiver. The cache set which has evictions during speculative execution will give longer access time.

\reffig{fig:eval_spectre_pp} shows the prime-probe Spectre v1 attack (accessing a secret of 30) manages to evict the receiver's circled cache line whose index is 30 in the SA-LRU system (blue). There are more than one peak of high access latency because there are other memory accesses in the sender code. These unrelated accesses cause conflicts at fixed cache sets while the circled peak moves as the secret value changes. The result shows while the prime-probe channel does not require shared memory regions, its result contains more noise than the flush-reload speculative execution attack.

The fully-associative cache of \cacheinvfa~(orange) defeats the attack as shown in \reffig{fig:eval_spectre_pp}. The sender code running in \cacheop~(green) will always cause a random eviction of cache lines upon either a \textit{Tag Miss} or a \textit{Mapping Miss}, which causes no secret-dependent timing difference.

While there are many other variants of speculative attacks, they mostly differ in how the secret is illegally e accessed. Since we are only interested in preventing the secret from being leaked out through a cache timing channel, the tests with flush-reload and prime-probe Spectre V1 attacks are representative. We also tested with the speculative store bypass attack \cite{spectressb} using both flush-reload and prime-probe covert channels, and obtained similar results (not shown).

%% file: 0800_performance_eval.tex
\subsection{Performance Evaluation}
\label{sec_perf_eval}

The performance of tested architectures is measured by running 24 benchmarks with the reference dataset in the SPEC CPU2006 benchmark suite. We skip the first 10 billion instructions in the benchmarks and run the next 500 million instructions. \cacheinvfa~has two tested configurations, \cacheinvfaone~with a 1-cycle fully-associative L1D cache and \cacheinvfatwo~with a 2-cycle L1D cache. \cacheop~has four tested configurations with 0, 2, 4 and 6 extra index bits, denoted as \cacheop-k0, k2, k4 and k6. \cacheop-k4 is shown and explained in \reffig{fig:cache_arch}.

\bheading{Overall performance.}
\reffig{fig:eval_perf} (a) demonstrates the relative performance of \cacheabbrev~in terms of execution time compared to the SA-LRU baseline whose performance is normalized to 1. The last column of \reffig{fig:eval_perf} (a) is the geometric mean of overhead.

The average execution time of \cacheinvfaone~is 5.6\% longer than SA-LRU. It improves the performance of zeusmp by 1.3\% in the best case, showing both better security and higher performance. The worst-case overhead is 39.6\% for mcf. Having a slower fully-associative L1 cache in \cacheinvfatwo~increases the average overhead to 10.8\%.

For \cacheop~(blues bars), adding additional index bits, \textit{k}, improves the performance. \cacheop-k0 has a high average overhead of 15.3\%. \cacheop-k2 reduces the overhead to 9.4\% (5.9\% better than \cacheop-k0). \cacheop-k4 has an overhead of 7.2\% (2.2\% better than \cacheop-k2) and \cacheop-k6 has an overhead of 6.8\% (0.4\% better than \cacheop-k4). Having more index bits improves the performance but the incremental benefit is decreasing. Further increasing $k$ may cause the cache access latency to increase, reducing overall system performance.

Since we use the same hardware configurations, we can compare \cacheinvfaone~(overhead 5.6\%) and \cacheop-k6 (overhead 6.8\%) to leading defenses such as InvisiSpec \cite{invisispec:correction} with 16.8\% overhead and CleanupSpec \cite{invisispec:correction} with 5.1\% overhead. However, InvisiSpec and CleanupSpec do not prevent all non-speculative side-channel attacks, which \cacheinvfa~and \cacheop~do.

We also see that \cacheinvfa~is better if a fast fully-associative cache design that has the same access latency as SA-LRU is feasible. Although \cacheinvfaone~has an overhead of 5.6\%, the overhead of \cacheinvfatwo~becomes 10.8\%. In this case, \cacheop-k6 which has an overhead of only 6.8\% becomes the better choice.

\bheading{Factors causing performance overhead.} There are two major factors which cause performance degradation in \cacheabbrev. The first one affecting both \cacheinvfa~and \cacheop~is squashed loads which need to send \sqinv~to invalidate cache lines upon incorrect speculation. The second one affecting only \cacheop~is the \textit{ForwardNoFill} path upon a mapping hit but tag miss (see \reffig{fig:access_policy}), which prevents the cache fill and performs a cache line eviction. Both can lead to a higher cache miss rate and therefore lower performance.

\reffig{fig:eval_perf} (b) shows the number of \sqinv~requests that are sent in each benchmark. A strong correlation is observed when we compare \reffig{fig:eval_perf} (a) with \reffig{fig:eval_perf} (b). Benchmarks incurring highest performance overhead in both \cacheinvfa~and \cacheop~systems, e.g., astar, bzip2, calculix and mcf, are also the benchmarks requiring more \sqinv~operations. This means that speculatively accessed cache lines are likely to be used by later instructions. While invalidating such cache lines degrades performance, this feature is needed to enforce the security against speculative execution attacks.

\reffig{fig:eval_perf} (c) shows the number of speculative loads which have a mapping hit but tag miss (TagMiss load). For TagMiss loads, cache fills are disallowed and data is directly forwarded to the processor; a random cache line eviction is also triggered (see the middle red path in \reffig{fig:access_policy}). 
The advantage of \cacheop~having more index bits can be clearly seen. With no extra bits, \cacheop-k0 has a large number of TagMiss loads in benchmarks such as leslie3d. By adding a few extra bits, we can significantly reduce the number of TagMiss loads.

We show that even using a small \textit{k} such as 2, 4 and 6 can greatly reduce these cache fill preventions and cache line evictions. In \reffig{fig:eval_perf} (c), \cacheop-k2, \cacheop-k4 and \cacheop-k6 reduce the number of TagMiss loads doing \textit{ForwardNoFill} by an average of 70.4\%, 91.5\% and 97.3\% respectively for all benchmarks. This is because having more bits in the index field can reduce the chance of having a mapping hit (C.Index == L.Index in \reffig{fig:access_policy}), allowing more useful replacements. Similar to the observations for overall performance, we recommend k = 4 or 6 as the best options.

\bheading{Fraction of wrong-path loads.} We collect the fraction of speculative loads that are squashed for all benchmarks. This ranges from 0.0\% to 52.7\%, with an average of 11.1\%. This means most of speculative loads are correct-path instructions, which is also why CleanupSpec and our proposed \sqinv~have lower performance overhead.

